# Effect of the Darrieus-Landau instability on turbulent flame velocity

Maxim Zaytsev[1, 2] and Vitaliy Bychkov[1]

[1]*Institute of Physics, Umeå University, S-901 87 Umeå, Sweden*

[2]*Moscow Institute of Physics and Technology, 141 700, Dolgoprudny, Russia*

## Abstract

Propagation of turbulent premixed flames influenced by the intrinsic hydrodynamic flame instability (the Darrieus-Landau instability) is considered in a two-dimensional case using the model nonlinear equation proposed recently [1]. The nonlinear equation takes into account both influence of external turbulence and intrinsic properties of a flame front, such as small but finite flame thickness and realistically large density variations across the flame front. Dependence of the flame velocity on the turbulent length scale, on the turbulent intensity and on the density variations is investigated in the case of weak non-linearity and weak external turbulence. It is shown that the Darrieus-Landau instability influences the flamelet velocity considerably. The obtained results are in agreement with experimental data on turbulent burning of moderate values of the Reynolds number.

**Communication address:**

V. Bychkov,

Inst. of Physics, Umeå University

S-901 87, Umeå, Sweden

tel. (46 90) 786 79 32 fax: (46 90) 786 66 73

**e-mail:** **vitaliy.bychkov@physics.umu.se**

## I. Introduction

Turbulent flame velocity $U_w$ is one of the basic characteristics of premixed turbulent combustion, it is a key parameter for phenomenological models of the combustion process and a promising tool for multi-dimensional computations of turbulent burning in real industrial devices [2, 3]. Turbulent combustion may proceed in several distinctive regimes with quite different properties [2]. Among them the so-called "flamelet" regime is the most typical for practical systems like gas turbines of power plants or car engines. In the case of flamelet burning a chemical reaction occurs at fast time-scales and short length-scales relative to the turbulent ones. A flamelet propagates as a relatively thin front with the inner structure similar to the laminar flame of some thickness $L_f$, but turbulent flow distorts the flame strongly on large length scales in comparison with $L_f$. Larger surface area of a corrugated flame front leads to larger consumption rate of the fuel mixture, larger total heat release and larger velocity of flame propagation $U_w$ in comparison with the laminar flame velocity $U_f$. As a result, turbulence increases power available from a turbine combustor or internal combustion engine.

The regime of turbulent flamelets has been studied theoretically for a long time and many interesting results have been obtained [4-9]. However, most of these results were restricted to the artificial limit of zero fuel expansion, when the density of the burnt matter $\rho_b$ is the same as in the fuel mixture $\rho_f$ with their ratio $\Theta = \rho_f / \rho_b = 1$. In this limit flame propagates passively in the turbulent flow without affecting the flow. Yet, most of the laboratory flames involve considerable thermal expansion $\Theta = 5 - 10$, for which flame interacts with the turbulent flow quite strongly. By this reason the artificial limit of zero thermal expansion $\Theta = 1$ cannot provide quantitative description of turbulent flames. Besides, thermal expansion leads to qualitatively new effects of flame-flow interaction such as the Darrieus-Landau (DL) instability. In the case of laminar flames the DL instability bends an initially planar flame front increasing the velocity of flame propagation [10-12]. Similar effects are expected for turbulent flamelets at least in the limit of weak and moderate turbulent intensity. The role of the DL instability for turbulent flamelets has been widely discussed, but in few papers one could find real attempts to solve the problem because of involved mathematical and computational difficulties [8, 13, 14]. However, even these papers considered only the limit of ultimately weak instability at small thermal expansion $\Theta - 1 << 1$ (or $\Theta < 2$ in [14]). If one takes the artificial limit of zero thermal expansion $\Theta = 1$ like [4-7,9], then the DL instability disappears. Even in the case of laminar flames the nonlinear theory of the DL instability has been restricted for a long time to the limit of small thermal expansion $\Theta - 1 << 1$ [15,16]. Quantitative nonlinear theoretical description of the DL instability for realistically large expansion factors $\Theta$ has been developed not so long ago [11,17,18]. Of course, the instability effect on turbulent flames could be studied by direct numerical simulations of the complete set of combustion equations, but simulations performed so far considered flame dynamics on small length scales below $10 L_f$ [19,20], for which the DL instability is thermally suppressed.

Recently a nonlinear equation has been proposed, which takes into account both influence of external turbulence and the DL instability with realistically large density variations across the flame front [1]. Preliminary evaluations of the turbulent flame velocity on the basis of the obtained equation [1,21] agree quite well with experimental results [22,23]. Still, the particular solutions to the nonlinear equation [22,23] did not include direct influence of the DL instability. Though such solution may be used to describe flame dynamics for a rather low integral turbulent length scale of about $(30 - 50)L_f$ like in experiments [23], a more common experimental situation involves large length scales exceeding $100L_f$ considerably [22], for which the DL instability affects flame velocity significantly. In the present paper we solve the nonlinear equation [1] in order to investigate the effect of the DL instability on dynamics of turbulent flames.

## II. Model equation

We solve the following model equation describing dynamics of a flame front $z = F(\mathbf{x}, t)$ in a weakly turbulent flow $\mathbf{u}(\mathbf{x}, t)$

$$\frac{\Theta + 1}{2\Theta}\left(1 + C_1 L_f \hat{\Phi}\right)\frac{\hat{\Phi}^{-1}}{U_f^2}\frac{\partial^2 F}{\partial t^2} + \left(1 + C_2 L_f \hat{\Phi}\right)\frac{1}{U_f}\frac{\partial F}{\partial t} + 1 - U_w/U_f + \frac{\Theta}{2}(\nabla F)^2 +$$

$$\frac{(\Theta - 1)^3}{16\Theta}\left[(\nabla F)^2 - \left(\hat{\Phi}F\right)^2\right] - \frac{\Theta - 1}{2}\left(1 - \frac{\lambda_c}{2\pi}\hat{\Phi}\right)\hat{\Phi}F - \left(1 + \frac{\hat{\Phi}^{-1}}{U_f}\frac{\partial}{\partial t}\right)\frac{u_z}{U_f} = 0. \quad (1)$$

The equation (1) has been proposed in [1] in the section below we explain the origin and the main components of the equation. Besides, we consider some particular solutions to Eq. (1) obtained before. Equation (1) is written in the reference frame of the average position of a statistically stationary turbulent flame front, with the average velocity $U_w$ of flame propagation being slightly different from the laminar flame velocity $U_f$. The main advantage of Eq. (1) in comparison with earlier models of turbulent flames [4-6] is that it takes into account realistically large density variations across the flame front described by the factor $\Theta = \rho_f / \rho_b$, which is the density ratio of the fuel mixture $\rho_f$ and the burnt matter $\rho_b$. Parameters $\lambda_c$, $C_1$, $C_2$ are related to internal thermal-chemical properties of the flame front: These parameters have been found in the linear theory of the DL instability of a flame of finite thickness $L_f$ [24,25]. The operator $\hat{\Phi}$ implies multiplication by absolute value of the wave number component along the flame surface in Fourier space, which may be presented in the case of a 2D flow as

$$\hat{\Phi}F = \frac{1}{2\pi}\int |k| F_k \exp(ikx)dk \qquad (2)$$

Model equation (1) has been proposed on the basis of three rigorous theories: the linear theory of the DL instability by Pelce and Clavin [24], the nonlinear theory of curved flames resulting from the instability by Bychkov [11] and the linear theory of flame response to weak turbulence by Searby and Clavin [26]. Below we explain briefly the main results of the theories as well as components of the model equation (1).

It has been shown [24] that development of small perturbations at an initially planar flame front may be described by the equation

$$\frac{\Theta+1}{2\Theta}\left(1+C_1 L_f \hat{\Phi}\right)\frac{\partial^2 F}{\partial t^2} + \left(1+C_2 L_f \hat{\Phi}\right)U_f \hat{\Phi}\frac{\partial F}{\partial t} - \frac{\Theta-1}{2}\left(1-\frac{\lambda_c}{2\pi}\hat{\Phi}\right)U_f^2 \hat{\Phi}^2 F = 0 \qquad (3)$$

Equation (3) has been obtained taking into account small but finite flame thickness $L_f$. The numerical factors $C_1$, $C_2$ and the cut-off wavelength $\lambda_c$ depend on internal flame structure [24,25]. Particularly, in the case of a flame with Lewis number equal unity and a constant coefficient of thermal conduction the respective formulas become

$$\lambda_c = 2\pi L_f\left(1+\Theta\frac{\Theta+1}{(\Theta-1)^2}\ln\Theta\right), \qquad C_1 = 0, \qquad C_2 = \frac{\Theta\ln\Theta}{\Theta-1}. \qquad (4)$$

In the case of an infinitely thin front Eq. (3) goes over to the Darrieus-Landau dispersion relation since $\lambda_c \propto L_f$ [24]. According to Eq. (3) small perturbations of a planar flame grow exponentially in time if the perturbation wavelength exeeds the cut-off wavelength $\lambda_c$. Equation (3) is the first component of the nonlinear model (1).

If a flame front propagates in a 2D channel of width $R$ with ideally slip and adiabatic walls, then the DL instability is thermally suppressed in narrow tubes with $R < R_c = \lambda_c / 2$. In wider tubes $R > R_c$ perturbations grow exponentially until nonlinear effects come to play. In tubes of a moderate width $R < (4-5)R_c$ nonlinear stabilization takes place leading to a smooth curved stationary flame shape [11,12,26]. The curved flame shape and velocity may be described by the nonlinear equation derived in [11]:

$$1 - U_w / U_f + \frac{\Theta}{2}(\nabla F)^2 + \frac{(\Theta-1)^3}{16\Theta}\left[(\nabla F)^2 - (\hat{\Phi}F)^2\right] - \frac{\Theta-1}{2}\left(1 - \frac{\lambda_c}{2\pi}\hat{\Phi}\right)\hat{\Phi}F = 0, \quad (5)$$

where $U_w$ is propagation velocity of a curved (wrinkled) flame front, which is larger than $U_f$. The nonlinear terms in Eq.(5) take into account both cusp formation at the flame front and vorticity production behind a curved flame. Equation (5) has been derived for arbitrary expansion coefficients $\Theta$ (even large ones) assuming small but finite flame thickness similar to the linear theory [24]. The thermal-chemical properties of the burning mixture are taken into account in Eq. (5) by the cut-off wavelength $\lambda_c$, the analytical formula for which coincides with the expression obtained in the linear theory. The combination of equations (3) and (5) determines the effects of flame-flow interaction in the equation (1) in absence of external turbulence. Influence of turbulence is taken into account according to the linear theory [26] for a flame front in a weak external flow

$$\frac{\Theta+1}{2\Theta}\frac{\partial^2 F}{\partial t^2} + U_f\hat{\Phi}\frac{\partial F}{\partial t} - \frac{\Theta-1}{2}U_f^2\hat{\Phi}^2 F - \left(U_f\hat{\Phi} + \frac{\partial}{\partial t}\right)u_z = 0, \quad (6)$$

where $u_z$ is the z-component of turbulent velocity at the surface $z = 0$. Assuming weak nonlinear effects, weak turbulence and a thin flame front we can combine (3), (5), (6) into one model equation (1), as it has been proposed in [1].

In the present paper we consider a 2D flame propagating in a "tube" of width $R$ with adiabatic boundary condition at the walls

$$\frac{\partial F}{\partial x} = 0 \qquad \text{at} \qquad x = 0, R,$$ (7)

so that the flame shape may be presented as

$$F = \sum F_n \cos\left(\frac{\pi n x}{R}\right)$$ (8)

with the operator $\hat{\Phi}$

$$\hat{\Phi} F = \sum \frac{\pi n}{R} F_n \cos\left(\frac{\pi n x}{R}\right).$$ (9)

Equation (1) for a 2D flow becomes

$$\frac{\Theta + 1}{2\Theta}\left(1 + C_1 L_f \hat{\Phi}\right)\frac{\hat{\Phi}^{-1}}{U_f^2}\frac{\partial^2 F}{\partial t^2} + \left(1 + C_2 L_f \hat{\Phi}\right)\frac{1}{U_f}\frac{\partial F}{\partial t} + 1 - U_w / U_f + \frac{\Theta}{2}\left(\frac{\partial F}{\partial x}\right)^2 +$$

$$\frac{(\Theta - 1)^3}{16\Theta}\left[\left(\frac{\partial F}{\partial x}\right)^2 - \left(\hat{\Phi} F\right)^2\right] - \frac{\Theta - 1}{2}\left(1 - \frac{R_c}{\pi}\hat{\Phi}\right)\hat{\Phi} F - \left(1 + \frac{\hat{\Phi}^{-1}}{U_f}\frac{\partial}{\partial t}\right)\frac{u_z}{U_f} = 0.$$ (10)

Incompressible 2D turbulence in the laboratory reference frame may be described by the representation [8]

$$u_z = \sum U_i \cos(k_i z + \varphi_i)\cos(k_i x),$$ (11)

$$u_x = \sum U_i \sin(k_i z + \varphi_i)\sin(k_i x)$$ (12)

where we have taken the continuity condition into account. In Eqs. (11), (12) $k_i = \pi i / R$ are the wave numbers of turbulent harmonics and $\varphi_i$ stand for random phases. The amplitudes $U_i$ of turbulent harmonics are determined by the Kolmogorov spectrum $U_i \propto k_i^{-5/6}$ with the rms-turbulent velocity in one direction given by the formula

$$U_{rms}^2 = \sum U_i^2 / 4.$$ (13)

The number of turbulent harmonics in (11), (12) is a free parameter of our model. In the case of a freely propagating flame front the x-dependence of turbulent velocity also includes random phases. However, since we are interested in flame propagation in a channel of finite width with slip walls, then the boundary condition of the walls requires zero phases. In general, the turbulent velocity field (11), (12) should also involve temporal pulsations. The influence of temporal pulsations has been investigated recently in [27], where it has been shown that temporal pulsations do not lead to any qualitatively new effect in turbulent flame propagation. By this reason in the present paper we will use the Taylor hypothesis of "stationary" turbulence, for which pulsations caused by flame propagation are much stronger than "real" temporal pulsations of the flow velocity in (11), (12). The Taylor hypothesis has been used in many papers on turbulent flame dynamics [4, 6-8, 28]. Since the model equation (1) is written in the reference frame of a statistically stationary turbulent flame, then we have to go over to the same reference frame in the formulas for the turbulent velocity field (11), (12). In the case of weak turbulence we have the longitudinal turbulent velocity field at the front position $z = U_f t$

$$u_z = \sum U_i \cos\left(U_f k_i t + \varphi_i\right)\cos\left(k_i x\right).$$ (14)

The equation (1) has been solved before in two particular cases:

A) no turbulence $U_{rms} = 0$, when dynamics of the flame front is affected by the DL instability only;

B) no direct influence of the DL instability, when only external turbulence controlls flame propagation.

Below we consider briefly solution to Eq.(1) in both cases.

### A) *Influence of the DL instability only*

If the turbulent intensity is zero $U_{rms} = 0$, then the model equation (1) describes linear development of the DL instability at an initially planar flame front and subsequent propagation of a wrinkled flame resulting from the instability. As it has been pointed above, the instability developes and a curved flame shape becomes possible in sufficiently wide tubes $R/R_c > 1$ with the critical tube width $R_c$ determined by thermal-chemical flame parameters: for example, $R_c = \lambda_c/2$ in the case of a 2D flow in a channel with ideally slip and adiabatical walls. If the channel width is not too large $R/R_c < 4 - 5$, then the DL instability results in a smooth curved flame shape propagating with velocity [11]

$$U_w/U_f - 1 = \frac{2\Theta(\Theta-1)^2}{\Theta^3 + \Theta^2 + 3\Theta - 1} M \frac{R_c}{R}\left(1 - M\frac{R_c}{R}\right), \qquad (15)$$

where $M = \text{Int}[2R/R_c + 1/2]$. Figure 1 presents dependence of the wrinkled flame velocity on the tube width $R/R_c$ for the expansion factors $\Theta = 5, 7, 9$. As one can see, the wrinkled flame velocity exceeds the planar flame velocity because of the larger surface of fuel consumption. The difference between the wrinkled and planar flame velocities becomes larger for larger thermal expansion $\Theta$, since larger $\Theta$ (larger density difference across the flame) leads to stronger DL instability. The analytical formula (15) agrees quite well with results of direct numerical simulations of flame dynamics in tubes [12, 29]. As the tube width increases $R/R_c \rightarrow \infty$ the wrinkled flame velocity tends to a limiting value

$$U_w/U_f - 1 = \frac{1}{2}\frac{\Theta(\Theta-1)^2}{\Theta^3 + \Theta^2 + 3\Theta - 1}, \qquad (16)$$

which determines a maximal possible velocity increase for a stationary wrinkled flame. However, the maximal velocity may be achieved even for tubes of a moderate width, for example, for $R = 2R_c$. The maximal velocity increase (16) depends only on the expansion factor $\Theta$.

In the present paper we are interested, mostly, in flame propagation in channels of moderate width $R/R_c < 4-5$, for which the DL instability results in stationary wrinkled flames. In much wider tubes the curved stationary shape becomes unstable with respect to the secondary DL instability of a small scale [17]. At that point it is interesting that the model equation (1) describes also the stability limits of the secondary DL instability, which are in good agreement with the results of direct numerical simulations [12].

### B) Influence of external turbulence only

The model equation (1) has also a particular solution related to the external turbulence only with no direct influence of the DL instability. Such solution has been found in [1] for a three-dimensional (3D) turbulent flow. We now obtain a 2D version of that solution taking into account the condition of weak turbulence. In the case of a turbulent flow (14) we can rewrite the turbulent terms of Eq. (1) as follows

$$\left(1 + \frac{\hat{\Phi}^{-1}}{U_f}\frac{\partial}{\partial t}\right)\frac{u_z}{U_f} = \sum \frac{U_i}{U_f}\Big[\cos\left(U_f k_i t + \phi_i\right) - \sin\left(U_f k_i t + \phi_i\right)\Big]\cos k_i x =$$

$$= \sum \sqrt{2}\frac{U_i}{U_f}\cos\left(U_f k_i t + \phi_i + \frac{\pi}{4}\right)\cos k_i x \qquad (17)$$

and look for a flame front position $F = F(x,t)$ in a similar form

$$F = \sum F_i(t)\cos(k_i x). \qquad (18)$$

Substituting (18) into the model equation (1) with the accuracy of linear terms we come to a system of ordinary differential equations for $F_i$

$$\frac{\Theta+1}{2\Theta}\left(1+C_1 L_f k_i\right)\frac{1}{k_i U_f^2}\frac{d^2 F_i}{dt^2}+\left(1+C_2 L_f k_i\right)\frac{1}{U_f}\frac{dF_i}{dt}-$$

$$\frac{\Theta-1}{2}\left(1-R_c k_i/\pi\right)k_i F_i=\sqrt{2}\frac{U_i}{U_f}\cos\left(U_f k_i t+\phi_i+\frac{\pi}{4}\right) \qquad (19)$$

with the following solution

$$F_i(t)=\frac{\sqrt{2}U_i}{D_i k_i U_f}\cos\left(U_f k_i t+\phi_i+\frac{\pi}{4}+\gamma_i\right), \qquad (20)$$

where

$$D_i=\left[\left(\frac{\Theta-1}{2}\left(1-R_c k_i/\pi\right)-\frac{\Theta+1}{2\Theta}\left(1+C_1 L_f k_i\right)\right)^2+\left(1+C_2 L_f k_i\right)^2\right]^{1/2} \qquad (21)$$

and the phase shift $\gamma_i$ is determined by the expressions

$$\cos\gamma_i=\frac{1}{D_i}\left(\frac{\Theta-1}{2}\left(1-R_c k_i/\pi\right)-\frac{\Theta+1}{2\Theta}\left(1+C_1 L_f k_i\right)\right), \qquad (22)$$

$$\sin\gamma_i=-\frac{1}{D_i}\left(1+C_2 L_f k_i\right). \qquad (23)$$

The average turbulent flame velocity is related to nonlinear terms of Eq.(1)

$$U_w/U_f-1=\frac{\Theta}{2}\left\langle\left(\nabla F\right)^2\right\rangle+\frac{(\Theta-1)^3}{16\Theta}\left\langle\left(\nabla F\right)^2-\left(\hat{\Phi}F\right)^2\right\rangle, \qquad (24)$$

where $\langle..\rangle$ denotes time and space averaging. Substituting representation (18) with amplitudes (20) into (24) we find the turbulent flame velocity in the case of no direct influence of the DL instability

$$U_w/U_f-1=\frac{\Theta}{4U_f^2}\sum\frac{U_i^2}{D_i^2}. \qquad (25)$$

In the artificial limit of zero thermal expansion $\Theta=1$ and $R/L_f\to\infty$ the obtained formula (25) agrees with the well-known Clavin-Williams formula [4] written for the 2D turbulence model (11), (12):

$$U_w / U_f - 1 = \frac{U_{rms}^2}{2U_f^2} . \qquad (26)$$

Both the velocity amplitudes $U_i$ and the factors $D_i$ in Eq. (25) depend on the wavenumbers $k_i$, which, in turn, are determined by the tube width $k_i = i\pi / R$. As a result, equation (25) specifies dependence of the turbulent flame velocity on the tube width $R$. In order to understand the obtained dependence better we consider first a simplified case of a turbulent flow (14) modelled by a single harmonic

$$u_z = U_1 \cos\left(U_f k_1 t\right) \cos\left(k_1 x\right) . \qquad (27)$$

Such a simplified assumption about a turbulent flow is often used in direct numerical simulations [9, 27, 28]. In the case of a single turbulent harmonic the expression for turbulent flame velocity becomes

$$\frac{U_w}{U_f} - 1 = \Theta \left[ \left( 1 + \pi C_2 \frac{L_f}{R} \right)^2 + \left( \frac{\Theta - 1}{2} \left( 1 - \frac{R_c}{R} \right) - \frac{\Theta + 1}{2\Theta} \left( 1 + \pi C_1 \frac{L_f}{R} \right) \right)^2 \right]^{-1} \frac{U_{rms}^2}{U_f^2} . \qquad (28)$$

The dependence (28) is persented in Fig. 2 for the case of unit Lewis number $Le = 1$ and a constant coefficient of thermal conduction, for which the cut-off wavelength $\lambda_c = 2R_c$ and the numerical factors $C_1$, $C_2$ are determined by the formula (4). The turbulent flame velocity is shown for different values of the expansion coefficient $\Theta = 5, 7, 9$ and for different turbulent intensity $U_{rms} / U_f = 0.2, 0.5, 1$. As one can see, the dependence of turbulent flame velocity on the tube width differs considerably from the previous case of a flame affected by the DL instability only. Now the critical tube width $R = R_c$ is not a point of sharp bifurcation any more, but instead one has a smooth resonance at $R \approx R_c$ [26]. In narrow tubes $R < R_c$ before the resonance a curved shape of a flame front is also possible and the flame velocity exceeds the planar flame velocity $U_w > U_f$. Still, the narrower the tube, the stronger stabilizing effects of thermal conduction and finite flame thickness. According to Eq. (28) at $R \to 0$ the flame velocity is equal to the planar flame velocity. For wider tubes $R > R_c$ after the resonance the turbulent flame velocity decreases too, though for very wide tubes $R \gg R_c$ it tends to a finite limit

$$U_w / U_f - 1 = \frac{4\Theta^3}{4\Theta^2 + \left(\Theta^2 - 2\Theta - 1\right)^2} \frac{U_{rms}^2}{U_f^2}. \qquad (29)$$

A curious point is that the limiting values for the turbulent flame velocity Eq. (29) decrease with increase of the expansion coefficient $\Theta$. One can see the same tendency in Fig. 2a for $\Theta = 5, 7, 9$ and a fixed turbulent intensity $U_{rms} / U_f = 0.5$. The decrease of turbulent flame velocity in the case of weak turbulence with no direct influence of the DL instability has been obtained already in [1]. Such tendency is opposite to the previous case of a wrinkled flame shape caused by the DL instability only, for which flame velocity increases with $\Theta$. Besides, this tendency is also opposite to the situation of strong turbulence [1]: strongly turbulent flames propagate faster for larger expansion factors $\Theta$. There is no physical explanation yet, why turbulent flame velocity depends on thermal expansion $\Theta$ in different ways for the cases of weak and strong turbulence. Finally, the dependence of turbulent flame velocity $U_w / U_f$ on the turbulent intensity $U_{rms} / U_f$ in the case of weak turbulence is the same as in the Clavin-Williams formula: $U_w / U_f - 1 \propto U_{rms}^2 / U_f^2$.

The situation of a large number of turbulent harmonics in Eq. (14) is more realistic than a single turbulent mode. Particularly, Figure 3 presents dependence of turbulent flame velocity on the tube width for 150 turbulent harmonics. As in Fig. 2 we take unit Lewis number $Le = 1$ and a constant coefficient of thermal conduction. The main difference between the plots in Figs. 2 and 3 is that the resonance is practically missing for multi-scaled turbulence of Fig. 3. For example, for the turbulent intensity $U_{rms} / U_f = 0.5$ and the expansion factor $\Theta = 5$ the turbulent flame velocity depends on the tube width in a quite monotonic way increasing from $U_f$ in narrow tubes $R << R_c$ to the limiting value Eq. (29) in wide tubes $R >> R_c$. For larger expansion factors and larger turbulent intensity the slight resonance may be still seen. The reduced effect of resonance at $R \approx R_c$ for multi-scaled turbulence may be easily understood, since now a considerable part of turbulent energy is spread between harmonics of a smaller scale, which have resonance at different points. Apart from the resonance effect, other tendencies of the turbulent flame velocity remain qualitatively the same for the multi-scaled turbulence and for a single turbulent harmonic.

## III. Flame under simultaneous action of turbulence and the DL instability.

### A) The method of solution

The main purpose of the present paper is to understand flame dynamics affected both by the DL instability and by external turbulence, that is, to find a more general solution to Eq. (1). One possible way would be to perform direct numerical simulations of Eq. (1). However, direct numerical simulations are typically characterised by a relatively low accuracy than a solution to an eigenvalue problem, and simulation results are more difficult for analysis. By this reason, we will solve the equation (1) as an eigenvalue problem in a semi-analytical way taking into account the conditions of weak turbulence $U_{rms}/U_f << 1$ and weak nonlinear effects $(\partial F / \partial x)^2 << 1$ used in the derivation of Eqs. (3), (5), (6). In that sense we would like to stress that in tubes of moderate width with no external turbulence the DL instability results in a stationary curved flame front (see Sec. IIA). On the contrary, if flame is affected by external turbulence only with no direct influence of the DL instability, then the solution to Eq. (1) is strongly oscillating, see Eq. (20), Sec. IIB. Then it would be reasonable to look for solution to (1) in the form of a combination of a stationary term $G(x)$ and a strongly oscillating term $H(x,t)$:

$$F = G(x) + H(x,t). \tag{30}$$

Substituting (30) into Eq (1) and taking time-average we come to the following equation:

$$1 - U_w/U_f + \frac{\Theta}{2}\left(\frac{\partial G}{\partial x}\right)^2 + \frac{(\Theta-1)^3}{16\Theta}\left[\left(\frac{\partial G}{\partial x}\right)^2 - \left(\hat{\Phi}G\right)^2\right] +$$

$$\frac{\Theta}{2}\left\langle\left(\frac{\partial H}{\partial x}\right)^2\right\rangle_t + \frac{(\Theta-1)^3}{16\Theta}\left\langle\left(\frac{\partial H}{\partial x}\right)^2 - \left(\hat{\Phi}H\right)^2\right\rangle_t - \frac{\Theta-1}{2}\left(1 - \frac{R_c}{\pi}\hat{\Phi}\right)\hat{\Phi}G = 0, \tag{31}$$

where $\langle...\rangle_t$ denotes time averaging. The obtained equation is similar to the stationary equation (5) describing the nonlinear stage of the DL instability with some correction terms produced by turbulence. Eliminating (31) out of (1) we come to the time-dependent equation for "turbulent" part of the solution:

$$\frac{\Theta+1}{2\Theta}\left(1 + C_1 L_f \hat{\Phi}\right)\frac{\hat{\Phi}^{-1}}{U_f^2}\frac{\partial^2 H}{\partial t^2} + \left(1 + C_2 L_f \hat{\Phi}\right)\frac{1}{U_f}\frac{\partial H}{\partial t} + \Theta\frac{\partial G}{\partial x}\frac{\partial H}{\partial x} +$$

$$\frac{(\Theta-1)^3}{8\Theta}\left[\frac{\partial G}{\partial x}\frac{\partial H}{\partial x} - \hat{\Phi}G\hat{\Phi}H\right] + \frac{(\Theta-1)^3}{16\Theta}\left[\left(\frac{\partial H}{\partial x}\right)^2 - \left(\hat{\Phi}H\right)^2 - \left\langle\left(\frac{\partial H}{\partial x}\right)^2 - \left(\hat{\Phi}H\right)^2\right\rangle_t\right] +$$

$$\frac{\Theta}{2}\left[\left(\frac{\partial H}{\partial x}\right)^2 - \left\langle\left(\frac{\partial H}{\partial x}\right)^2\right\rangle_t\right] + \frac{\Theta-1}{2}\left(1 - \frac{R_c}{\pi}\hat{\Phi}\right)\hat{\Phi}H - \left(1 + \frac{\hat{\Phi}^{-1}}{U_f}\frac{\partial}{\partial t}\right)\frac{u_c}{U_f} = 0. \tag{32}$$

In the limit of weak turbulence and weak nonlinear effects all nonlinear terms in Eq. (32) may be neglected, since they give only small corrections to $H$, and we come to the linear equation:

$$\frac{\Theta+1}{2\Theta}\left(1 + C_1 L_f \hat{\Phi}\right)\frac{\hat{\Phi}^{-1}}{U_f^2}\frac{\partial^2 H}{\partial t^2} + \left(1 + C_2 L_f \hat{\Phi}\right)\frac{1}{U_f}\frac{\partial H}{\partial t} +$$

$$\frac{\Theta-1}{2}\left(1-\frac{R_c}{\pi}\hat{\Phi}\right)\hat{\Phi}H-\left(1+\frac{\hat{\Phi}^{-1}}{U_f}\frac{\partial}{\partial t}\right)\frac{u_z}{U_f}=0 \ . \tag{33}$$

Then the oscillating term $H(x,t)$ is specified by external turbulence $u_z$ only independent of the stationary term $G(x)$ (see Eq. (33)). On the other hand, averaged terms with $H(x,t)$ play the role of "external force" in the stationary equation (31). Solution to Eq. (33) coincides with the solution to Eq. (1) presented in Sec. IIB. Taking $N_T$ turbulent harmonics in the representation (14) we look for the oscillating term $H(x,t)$ in the form

$$H(x,t)=\sum_{i=1}^{N_T}H_i(t)\cos\left(\frac{\pi i x}{R}\right) \tag{34}$$

and find

$$H_i(t)=\frac{\sqrt{2}U_i}{D_i k_i U_f}\cos\left(U_f k_i t+\phi_i+\pi/4+\gamma_i\right) \tag{35}$$

where $D_i$ and $\gamma_i$ are determined by Eqs. (21) - (23). Formulas (34), (35) specify the "external force"

$$\frac{\Theta}{2}\left\langle\left(\frac{\partial H}{\partial x}\right)^2\right\rangle_t+\frac{(\Theta-1)^3}{16\Theta}\left\langle\left(\frac{\partial H}{\partial x}\right)^2-\left(\hat{\Phi}H\right)^2\right\rangle_t \tag{36}$$

in Eq. (31).

We solve Eq. (31) numerically with the boundary conditions (7) taking $G(x)$ in the form:

$$G(x)=\sum_{i=1}^{N}\frac{R}{\pi i}G_i\cos\left(\frac{\pi i x}{R}\right). \tag{37}$$

We would like to stress that the number of harmonics $N_T$ in Eq. (34) and $N$ in Eq. (37) are two different values. The former, $N_T$, shows the number of turbulent modes in the representation Eq. (14) and plays the role of a free parameter of the model.

Particularly, Eq. (28) has been obtained under the assumption of a single turbulent mode, $N_T = 1$. On the contrary, the value $N$ in Eq. (37) shows the number of Fourier harmonics in the numerical solution to Eq. (31), which is determined by the accuracy requirements.

In order to simplify the numerical solution it is useful to perform the following auxiliary calculations:

$$\left(\frac{\partial G}{\partial x}\right)^2 = -\frac{1}{2}\sum_{m=0}^{2N} A_m \cos\left(\frac{\pi m x}{R}\right) + \frac{1}{2}\sum_{l=-N}^{N} B_l \cos\left(\frac{\pi l x}{R}\right), \tag{38}$$

$$\left(\hat{\Phi}G\right)^2 = \left(\sum_{i=1}^{N} G_i \cos\left(\frac{\pi i x}{R}\right)\right)^2 = \frac{1}{2}\sum_{m=0}^{2N} A_m \cos\left(\frac{\pi m x}{R}\right) + \frac{1}{2}\sum_{l=-N}^{N} B_l \cos\left(\frac{\pi l x}{R}\right) \tag{39}$$

$$\left(\frac{\partial G}{\partial x}\right)^2 - \left(\hat{\Phi}G\right)^2 = -\sum_{m=0}^{2N} A_m \cos\left(\frac{\pi m x}{R}\right), \tag{40}$$

where $A_0 = A_1 = 0$, $A_m = \sum_{i=1}^{m-1} G_i G_{m-i}$ for $m \geq 2$, $B_0 = \sum_{i=1}^{N} G_i^2$, $B_l = \sum_{i=1}^{N-l} G_i G_{m+i}$ for $1 \leq l \leq N-1$ and $B_N = 0$. After substituting (34), (37) in (31) and averaging in time we obtain with help of (38)-(40) the following system of algebraic equations for $G_i$ and $U_w/U_f - 1$:

$$U_w/U_f - 1 = \frac{\Theta}{4}\left(B_0 + \frac{1}{U_f^2}\sum_{i=1}^{N_T} U_i^2/D_i^2\right), \tag{41}$$

$$G_i(1 - iR_c/R) - \frac{\Theta B_i}{\Theta - 1} + \frac{1}{2}\left(\frac{\Theta}{\Theta - 1} + \frac{(\Theta - 1)^2}{4\Theta}\right)\left(A_i + A_i^{turb}\right) = 0 \quad \text{for } i \geq 1, \tag{42}$$

where we have introduced the designation

$$A_i^{turb} = \begin{cases} D_{i/2}^{-2} U_{i/2}^2/U_f^2, & \text{if } i \bmod 2 = 0 \text{ and } 2N_T \geq i; \\ \\ 0, & \text{if } i \bmod 2 \neq 0 \text{ or } 2N_T < i. \end{cases} \tag{43}$$

The system (41), (42) has been solved numerically. However, instead of solving (42) directly we have introduced and solved the following system of ordinary differential equations involving virtual "time" $\xi$

$$\frac{dG_i}{d\xi} = G_i\left(1 - \frac{R_c}{R}i\right) - \frac{\Theta B_i}{\Theta - 1} + \frac{1}{2}\left(\frac{\Theta}{\Theta - 1} + \frac{(\Theta - 1)^2}{4\Theta}\right)\left(A_i + A_i^{turb}\right) \quad \text{for } i \geq 1 \qquad (44)$$

with arbitrary initial values but a fixed value of $R_c/R$. We have found numerically that solution to the system (44) tends to a unique set of "stationary" values $G_i$ with a sufficiently small step of virtual "time" $\xi$ independent of the chosen initial values of $G_i$. The obtained set of "stationary" values gives the solution to Eq. (42) and the turbulent flame velocity (41). The spectral method of numerical solution to boundary-value problem (31)-(7) described above, as it is well-known fact, provides the best accuracy among other methods. For example, even taking $N = 30$ Fourier harmonics in Eq. (37) we got 5% accuracy of calculations for the turbulent velocity $U_w/U_f - 1$. In most of the calculations we used a much larger value $N = 150$ providing the accuracy far better than 1% when the number of harmonics $N$ is doubled, and the Fourier expansion Eq. (37) converged well. As an illustration Fig. 4 shows spectrum of the numerically obtained solution of Eq. (42) for $R/R_c = 5$, $U_{rms}/U_f = 0.5$, $\Theta = 7$, $N = 150$ and $N_T = 150$. As one can see, in that case the amplitude of the first harmonic exceeds the amplitude of the 100th harmonics by 5 orders of magnitude. The chosen accuracy of numerical calculations was quite sufficient taking into account that assumptions used in basic equation (1) (that is, in the derivation of the original theories (3), (5), (6)) hold with much worth accuracy of about 30% and less. As a test of the numerical method we have solved first the stationary nonlinear equation (5), and the numerical result coinsided with the exact analytical solution Eq. (15). In order to imitate a multi-scaled turbulence we took the same number of turbulent harmonics in Eq. (34), $N_T = 150$. Besides, we have also investigated the case of a single turbulent mode $N_T = 1$, which may be useful in the future for comparison with direct numerical simulations.

### B) Results of numerical solution

The main dimensionless parameters that determine dynamics of curved flames in ideal tubes are the scaled tube width $R/R_c$, the turbulent intensity $U_{rms}/U_f$ and the expansion factor of the burning mixture $\Theta$. To be particular, in all calculations we took unite Lewis number $Le = 1$ and a constant coefficient of thermal conduction. First of all we have investigated dependence of the turbulent flame velocity $U_w/U_f - 1$ on the scaled tube width for different values of $\Theta$ and $U_{rms}/U_f$. We have been interested mostly in tubes of moderate width $R/R_c \leq 5$, realistic expansion coefficients $\Theta = 5 - 10$ and weak external turbulence $U_{rms}/U_f < 1$, for which the basic equation (1) may be valid.

The results of numerical solution are presented in Figs. 5-7 for $N_T = 150$ turbulent harmonics. Particularly, Fig. 5 shows scaled turbulent flame velocity $U_w/U_f - 1$ versus the scaled tube width $R/R_c$ for some fixed expansion factor ($\Theta = 5$, 7, 9 for Figs. 5(a), 5(b), 5(c) respectively) and different turbulent intensity. As one can see, the dependence remains qualitatively the same, as we had in the case of no influence of the DL instability, Sec. IIB, Fig. 3(b). In all plots turbulent flame velocity increases from the planar flame velocity $U_f$ in narrow tubes $R/R_c << 1$ to some limiting value in wide tubes $R/R_c >> 1$. The increase may be quite monotonic, as we have, for example, for $\Theta = 5$ and $U_{rms}/U_f = 0.5$. For smaller turbulent intensity $U_{rms}/U_f = 0.2$ one can easily see the trace of bifurcation "humps" typical for the case of the DL instability without turbulence, see Sec. IIA, Fig. 1. For larger turbulent intensity $U_{rms}/U_f = 1$ and larger expansion factors $\Theta = 7, 9$ the bifurcation "humps" vanish, but instead one can observe the resonance at $R \approx R_c$ described in Sec. IIb, Fig. 3. Judging by the qualitative look of the plots in Fig. 5, one could conclude that the influence of the DL instability becomes small already for $U_{rms}/U_f = 0.5$. However, this conclusion is wrong as one can see comparing Fig. 5(b) and Fig. 3(b) quantitatively. Indeed, choosing the problem parameters $R/R_c = 5$, $U_{rms}/U_f = 0.5$ and $\Theta = 7$ as an example we find increase of the turbulent flame velocity $U_w/U_f - 1 \approx 1.13$ taking into account the DL instabillity and $U_w/U_f - 1 \approx 0.42$ without the instability. Thus, though the DL instability does not bring anything qualitatively new into the plot, it leads to considerable quantitative changes increasing the scaled flame velocity almost 3 times! As a matter of fact, for such strong velocity increase the basic equation (1) does not hold any more, still it may show general features of flame front dynamics under simultaneous action of external turbulence and the DL instability. To make the difference between the cases with and without the DL instability more distinctive we present the characteristic velocity increase $U_w/U_f - 1$ at $R = 5R_c$ versus $U_{rms}/U_f$ in Fig. 6 for both cases. As one can see, the DL instability increases the turbulent flame velocity considerably for all turbulent intensities in the domain $U_{rms}/U_f < 1$. Besides, when turbulent intensity goes to zero $U_{rms}/U_f \to 0$, the instability still provides a non-zero increase of the flame velocity $U_w/U_f - 1 = 0.3 - 0.4$. Obviously, this does not happen for a particular solution without the instability. A curious point of Fig. 6 is that the obtained instability influence is much stronger than just formal adding of the "instability" solution of Sec. IIA to the "purely turbulent" solution of Sec. IIB.

Figure 7 presents plots similar to Fig. 5, but now we fix turbulent intensity and study the scaled velocity of flame propagation $U_w/U_f - 1$ for various expansion factors $\Theta$. At that point one has to remember, that $U_w/U_f - 1$ depends on $\Theta$ in a different way for the cases of a flame affected by the DL instability only or by weak external turbulence only. For the former case it increases with the expansion factor $\Theta$ (see Sec. IIA), while for the latter case it decreases (see Sec. IIB). One can observe both tendencies in Fig. 7 for the limiting values of $U_w/U_f - 1$ achived for relatively wide tubes, e.g. $R/R_c = 5$. In Fig. 7(a) the turbulent intensity is rather small $U_{rms}/U_f = 0.2$ and the characteristic flame velocity at $R/R_c = 5$ increases with $\Theta$ similar to the nonlinear stage of the DL instability. In Figs. 7(b)-(d) the turbulent intensity becomes larger $U_{rms}/U_f = 0.5 - 1$ and the flame velocity at $R/R_c = 5$ decreases with the expansion factor $\Theta$ as it happens for a weak turbulence with no direct influence of the DL instability.

We have also investigated the case of a single turbulent harmonic $N_T = 1$ popular in numerical simulations [9,27]. Figure 8 presents velocity of flame propagation for that case versus the scaled tube width for $\Theta = 5, 7, 9$ and $U_{rms}/U_f = 0.2, 0.5, 1$. The main difference between the cases of a single turbulent harmonic and a multi-scaled turbulence is a much more pronounced resonance at $R \approx R_c$. Figure 9 shows how velocity of flame propagation depends on the number of turbulent harmonics $N_T$: the more harmonics we take, the smoother resonance we get at $R \approx R_c$.

## IV. Discussion

The results on turbulent flame dynamics obtained in the present paper may be interpreted in two ways. First, they may be used to understand flame dynamics in experiments with a relatively small integral turbulent length scale (comparable to the cut-off wavelength of the DL instability $\lambda_c$) and relatively low turbulence intensity. Experiments of this type have been presented in [23], where the ratio of the channel width to the flame thickness (the Peclet number) varied as $R/L_f = 50 - 250$. Since the cut-off wavelenth is typically $\lambda_c = (20 - 50)L_f$ [24,25], then the experiment conditions in [23] are similar to the case of tubes of a moderate width $1 < R/R_c < 5$ considered in the present paper. Unfortunately, one cannot compare results of the present paper to the experimental results directly, since the present results describe a 2D flame, while the experimental flow is obviously a 3D one. Still, an indirect comparison is possible. For that purpose, one may remember that in the case of the DL instability the increase of flame velocity is twice larger for a 3D geometry in comparison with a 2D one [30]. The same relation holds for velocity increase of a weakly turbulent flame with no influence of the DL instability [1]. Thus, it would be quite natural to expect that, when both effects work together, the velocity increase is also twice larger for a 3D flow in comparison with a 2D flow of the present paper. Such evaluation is shown in Fig. 10 for $\Theta = 7$, $R/R_c = 5$ by the dashed line A, which agrees rather well with the experimental results [23] presented by markers. Accurate investigation of the turbulent flame velocity in a 3D flow requires time-consuming calculations and will be a subject of future work. For comparison, Fig. 10 shows also turbulent velocity for a 3D particular solution to Eq. (1) without direct influence of the DL instability (dashed line B). This plot goes noticeably below the experimental points, which demonstrates once more importance of the DL instability for turbulent flames.

Another possible interpretation of the present results is to treat them as a sub-grid study of flame dynamics in a combustor of a large scale with some turbulent intensity, which is not necessarily weak. In that case the tube width $R$ considered in the present paper should be interpreted as the upper limit of the short-wavelength band of the turbulent spectrum, and $U_{rms}$ plays the role of a turbulent velocity on a length scale $R$. Designating the "real" integral length scale and the "real" turbulent velocity by $L_T$ and $U_T$ we have for the Kolmogorov spectrum

$$U_{rms}/U_T = (R/L_T)^{1/3}. \tag{45}$$

The limit of weak turbulence implies $U_{rms}/U_f << 1$, and we took in our calculations $U_{rms}/U_f = 0.2 - 1$. Taking into account the evaluation for the cut-off wavelength [24,25] we find that the tubes of moderate width $R < 5R_c$ considered in the present paper correspond approximately to $R \approx 100 L_f$, or $R \approx 100\nu/U_f$, where $\nu$ is kinematic viscosity with a characteristic value $\nu \approx 0.15\,\text{cm}^2/\text{s}$. Then the integral velocity needed to produce the sub-grid turbulence with $U_{rms}/U_f = 0.2 - 1$ on the length scale $R \approx 100 L_f$ may be evaluated as

$$U_T/U_f = \frac{U_{rms}}{U_f}(R/L_T)^{-1/3} = \frac{U_{rms}}{U_f}\left(\frac{L_T U_f}{100\nu}\right)^{1/3}. \tag{46}$$

If we take typical laminar flame velocity $U_f = 100\,\text{cm/s}$ and the length scale $L_T = 100\,\text{cm}$ comparable to the size of a gas turbine, then calculations of the present paper correspond to the integral turbulent intensity $U_T/U_f = 1 - 10$ and to the respective turbulent Reynolds number $\text{Re}_T = U_T L_T/\nu = 10^5 - 10^6$. The obtained estimate shows that calculations of the present paper may be quite reasonable as a sub-grid model for a turbine combustor. Still in most of laboratory combustion experiments [3,22,23] a smaller integral turbulent length scale (about 1 cm in [23]) and a smaller value of the Reynolds number $\text{Re}_T = 10^2 - 10^4$ are employed.

**Acknowledgements**

This work has been supported by the Swedish Research Council (VR).

**Figures**

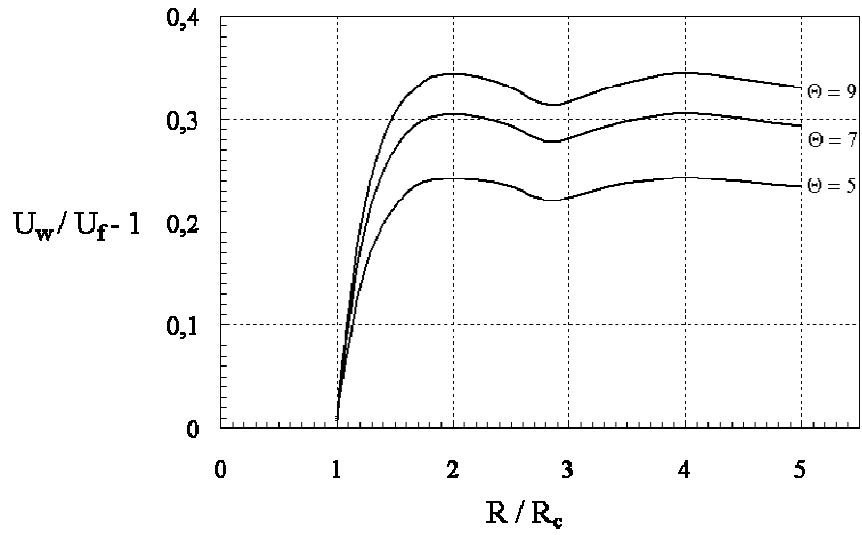

FIG. 1.   Scaled velocity of a curved stationary flame $U_w/U_f - 1$ vs the scaled tube width $R/R_c$ for different fuel expansion $\Theta = 5, 7, 9$ according to Eq. (15).

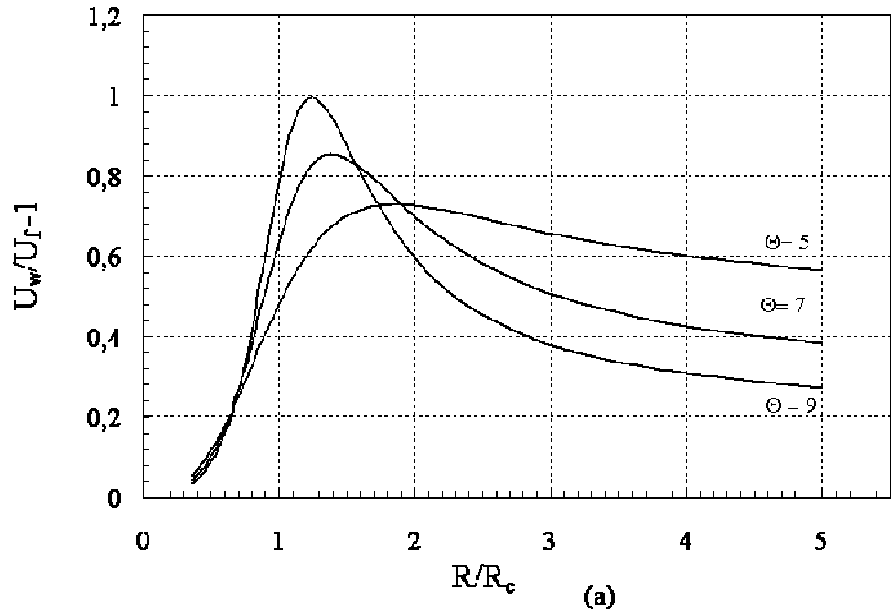

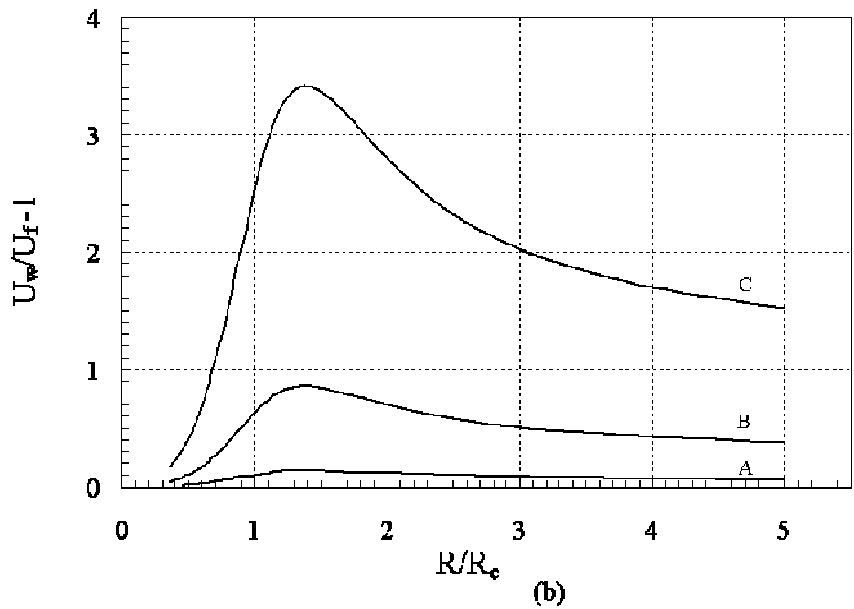

FIG. 2.   Scaled turbulent flame velocity $U_w/U_f - 1$ vs the scaled tube width $R/R_c$ given by the solution of Eq. (1), not including the DL instability, with one turbulent harmonic in representation (14): (a) for the fixed turbulent intensity $U_{rms}/U_f = 0.5$ but for different fuel expansion $\Theta = 5,7,9$; (b) for the fixed fuel expansion $\Theta = 7$ and different turbulent intensities $U_{rms}/U_f$: curves A, B, C correspond to values

$U_{rms}/U_f = 0.2, 0.5, 1$ respectively.

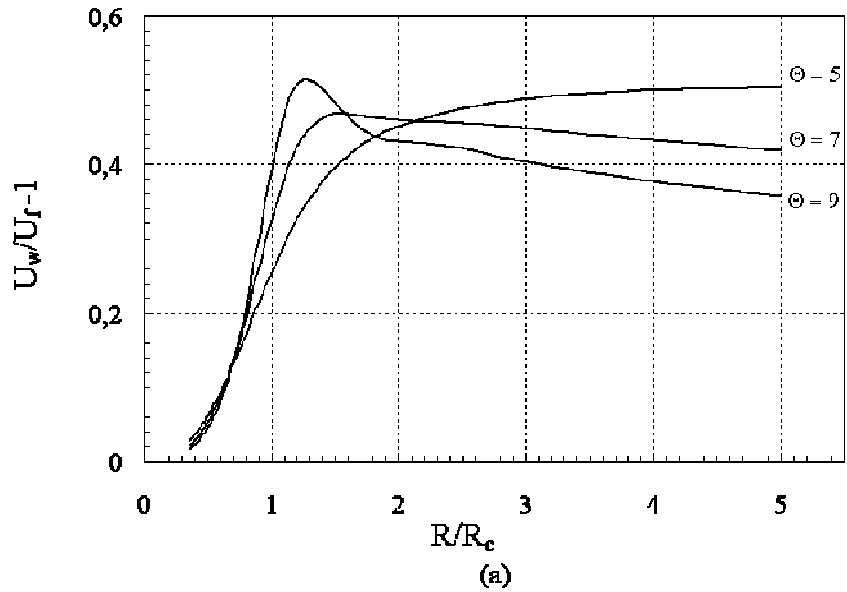

FIG. 3    Maxim Zaytsev    PHYSICAL REVIEW E

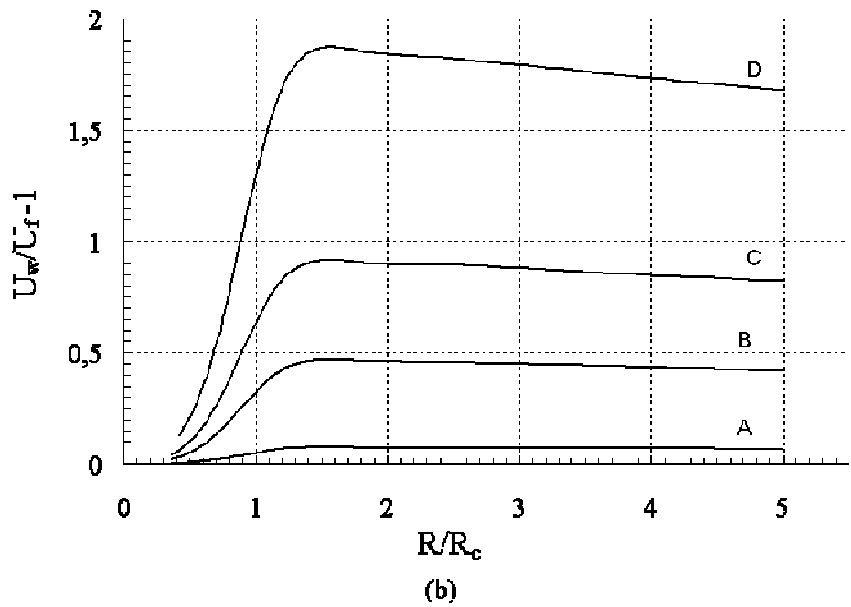

FIG. 3    Maxim Zaytsev    PHYSICAL REVIEW E

FIG. 3.    Scaled turbulent flame velocity $U_w/U_f - 1$ vs the scaled tube width $R/R_c$

given by the solution of Eq. (1), not including the DL instability, with 150 turbulent harmonics in representation (14): (a) for the fixed turbulent intensity $U_{rms}/U_f = 0.5$ but for different fuel expansion factors $\Theta = 5,7,9$; (b) for the fixed fuel expansion $\Theta = 7$ and different turbulent intensities $U_{rms}/U_f$: curves A, B, C, D correspond to values $U_{rms}/U_f = 0.2, 0.5, 0.7, 1$ respectively.

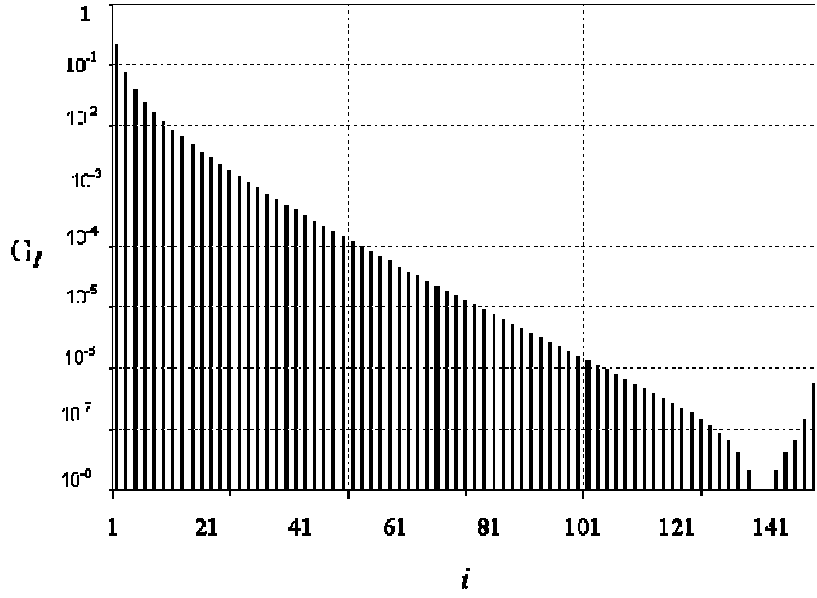

FIG. 4. Spectrum $G_i$ of the numerically obtained stationary solution $G(x)$ of Eq. (42) for $R/R_c = 5$, $U_{rms}/U_f = 0.5$, $\Theta = 7$, $N = 150$ and $N_T = 150$.

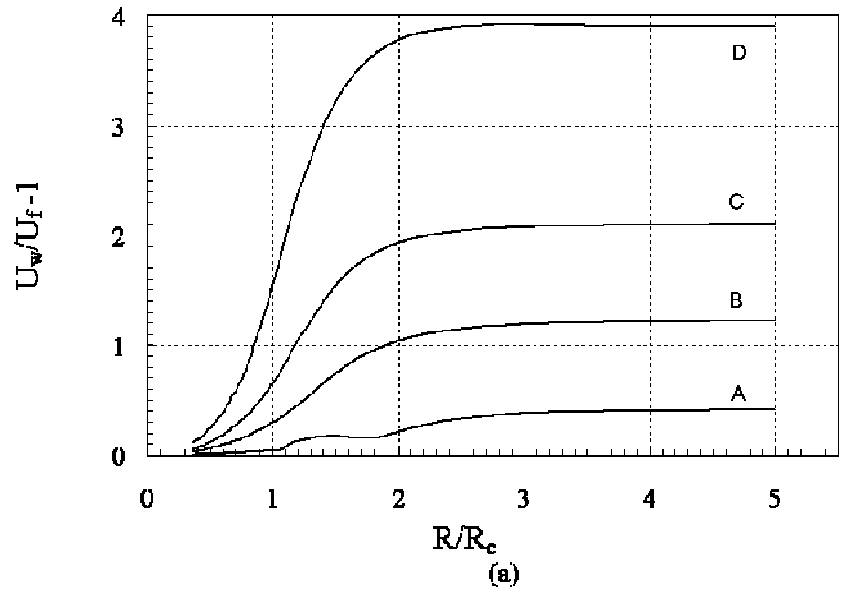

(a)

FIG. 5    Maxim Zaytsev    PHYSICAL REVIEW E

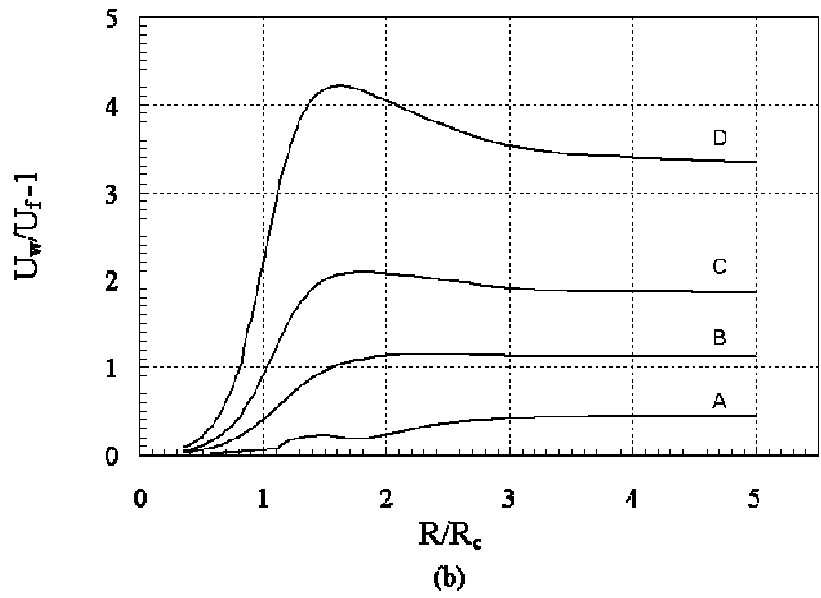

(b)

FIG. 5    Maxim Zaytsev    PHYSICAL REVIEW E

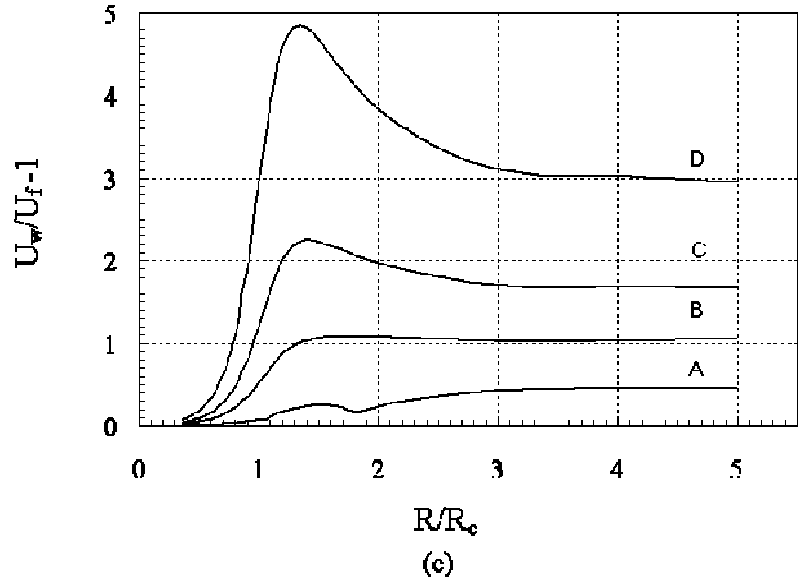

(c)

FIG. 5. Scaled turbulent flame velocity $U_w/U_f - 1$ versus the scaled tube width $R/R_c$ given by the solution of Eq. (1), including the DL instability, for some fixed fuel expansion factor ($\Theta = 5, 7, 9$ for Figs. 5(a), 5(b), 5(c) respectively) and different turbulent intensity $U_{rms}/U_f$: curves A, B, C, D correspond to values $U_{rms}/U_f = 0.2, 0.5, 0.7, 1$ respectively.

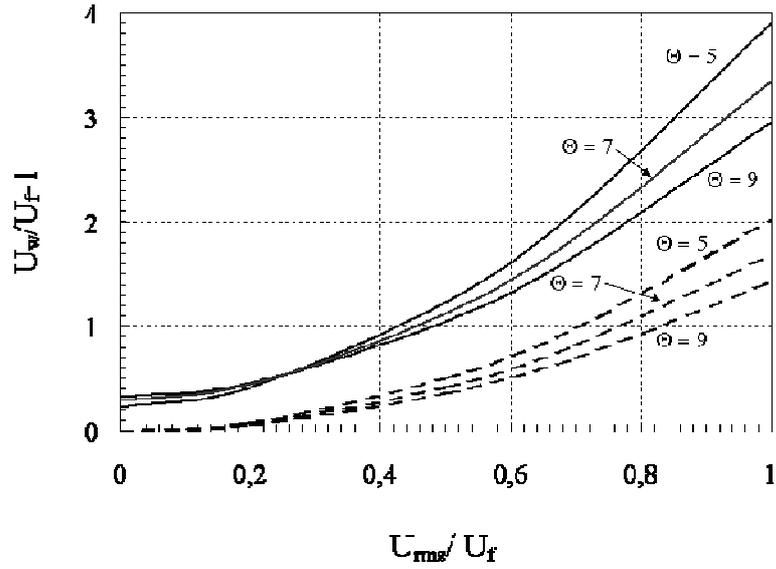

FIG. 6. Characteristic increase of the turbulent flame velocity $U_w/U_f - 1$ for $R = 5R_c$ versus turbulent intensity $U_{rms}/U_f$. The solid and dashed lines show the solutions with and without the DL instability, respectively.

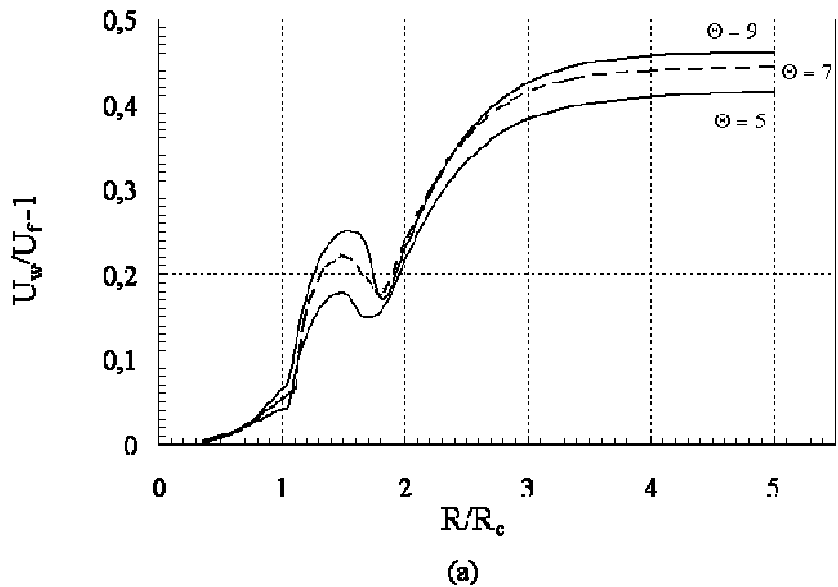

(a)

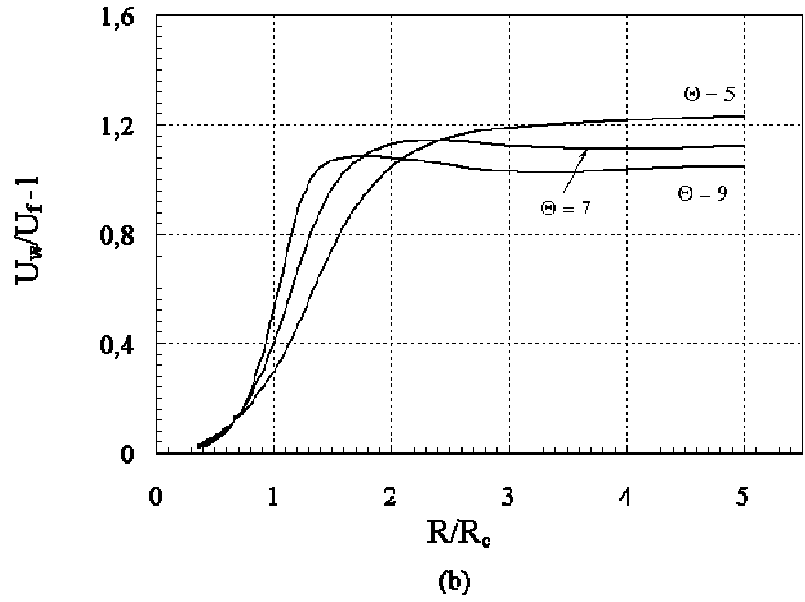

(b)

FIG. 7    Maxim Zaytsev    PHYSICAL REVIEW E

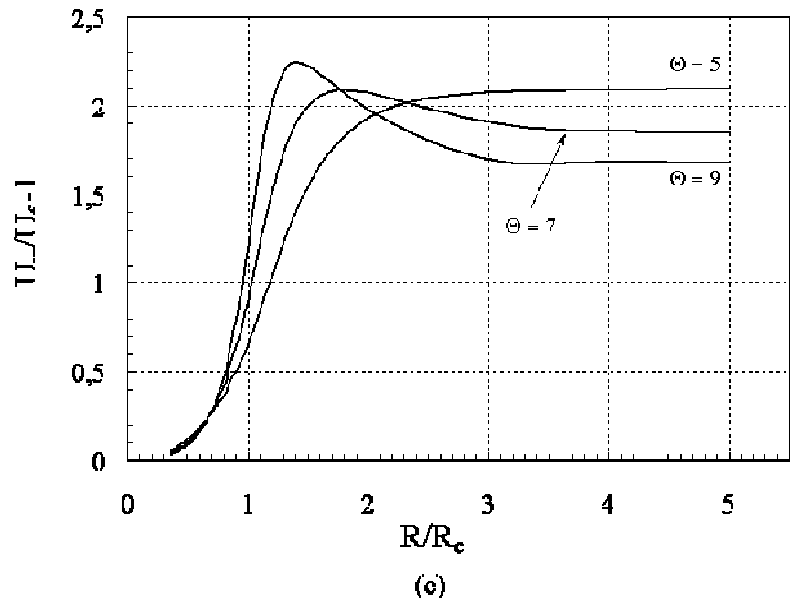

(c)

FIG. 7    Maxim Zaytsev    PHYSICAL REVIEW E

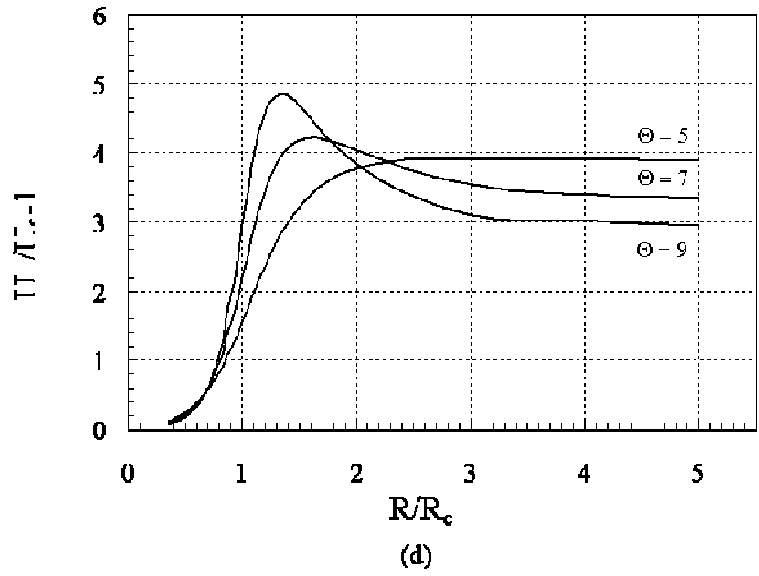

FIG. 7.  Scaled turbulent flame velocity $U_w / U_f - 1$ versus the scaled tube width $R / R_c$ given by the solution of Eq. (1), including the DL instability, for some fixed turbulent intensity ($U_{rms} / U_f = 0.2, 0.5, 0.7, 1$ for Figs. 7(a), 7(b), 7(c) 7(d) respectively) and different expansion factors $\Theta$. To distinguish plots we used dashed line in Fig. 7(a) for $\Theta = 7$.

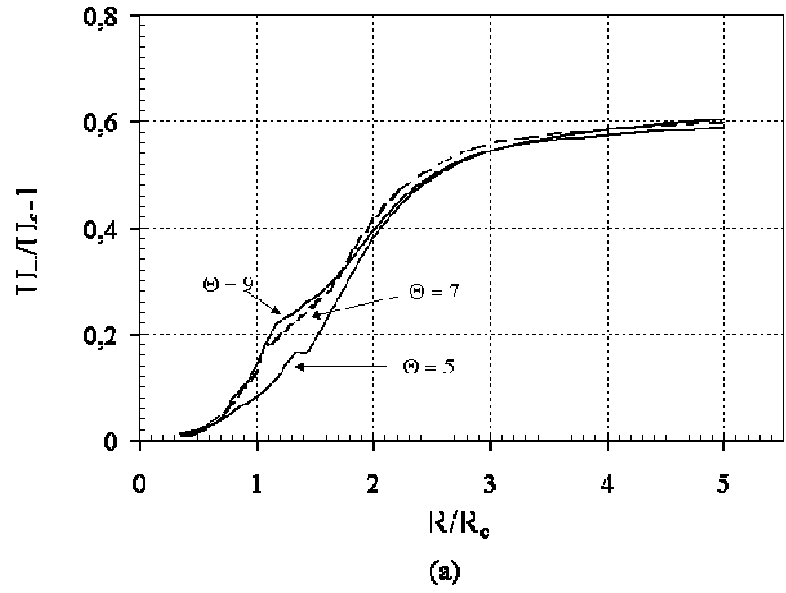

(a)

FIG. 8    Maxim Zaytsev    PHYSICAL REVIEW E

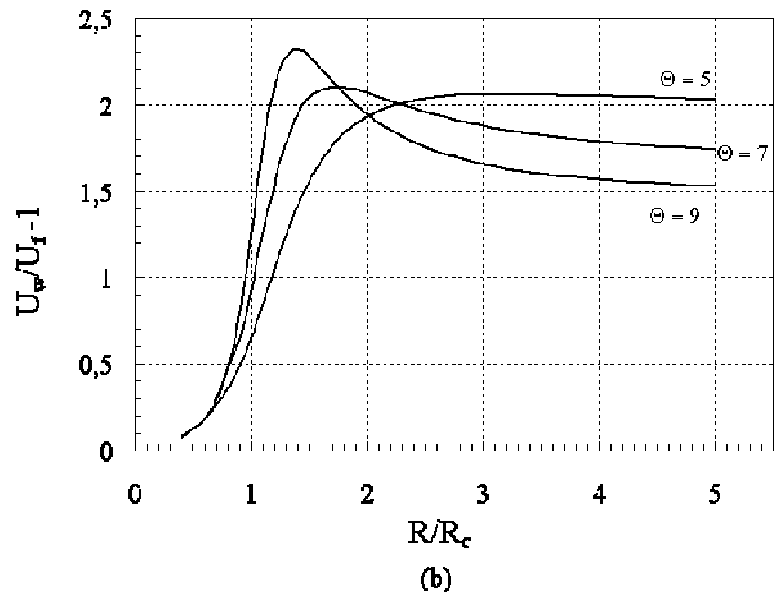

(b)

FIG. 8    Maxim Zaytsev    PHYSICAL REVIEW E

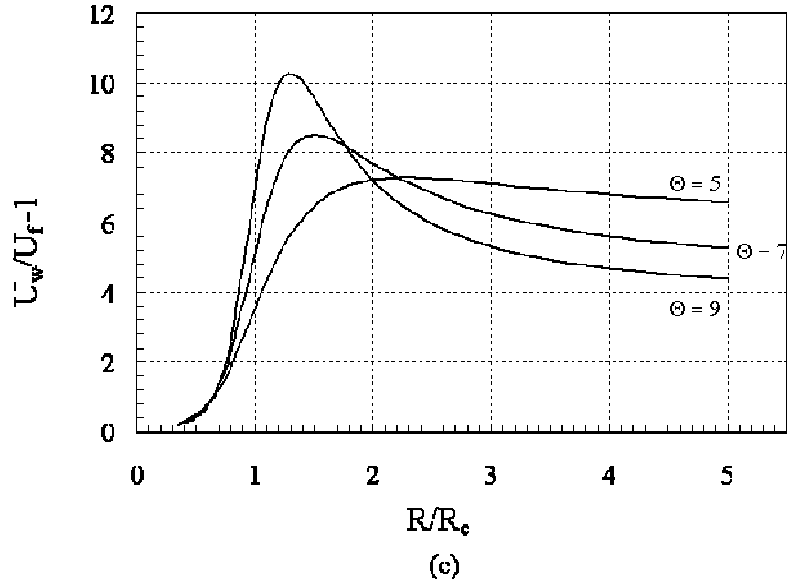

(c)

FIG. 8.    Scaled turbulent flame velocity $U_w / U_f - 1$ versus the scaled tube width $R / R_c$ given by the solution of Eq. (1), including the DL instability, with one turbulent harmonic in representation (14) for some fixed turbulent intensity ($U_{rms} / U_f = 0.2, 0.5, 1$ for Figs. 8(a), 8(b), 8(c) respectively) and different expansion factors $\Theta$.

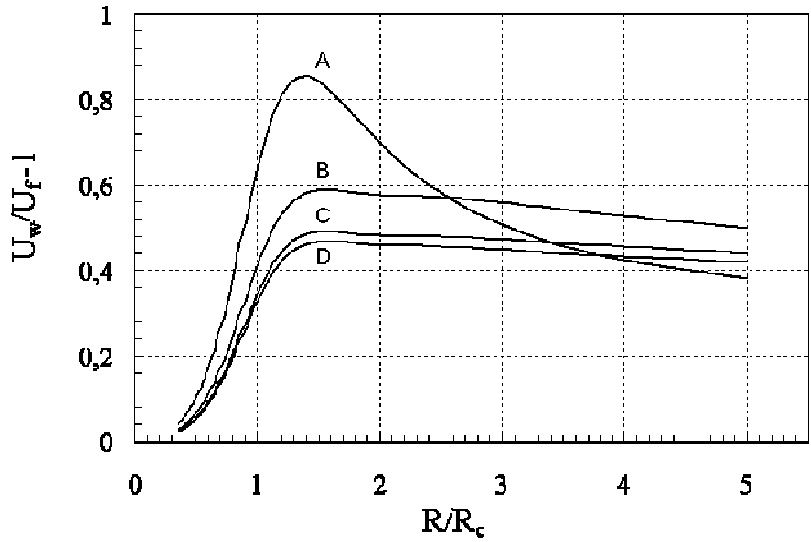

FIG. 9.   Scaled turbulent flame velocity $U_w/U_f - 1$ vs the scaled tube width $R/R_c$ given by the solution of Eq. (1), including the DL instability for different number of turbulent harmonics $N_T$ (curves A, B, C, D correspond to values $N_T = 1, 5, 30, 150$ respectively) and the fixed fuel expansion factor $\Theta = 7$ with $U_{rms}/U_f = 0.5$ and $N = 150$.

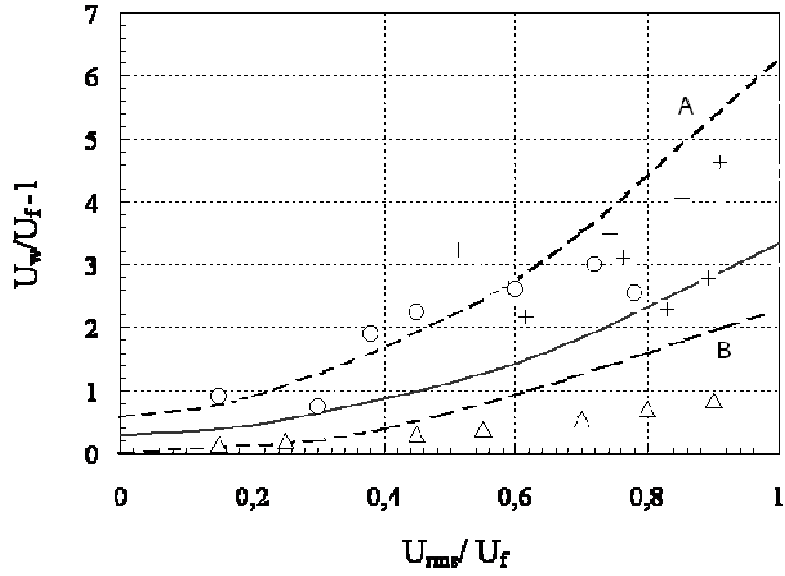

FIG. 10.    Scaled velocity of turbulent flame $U_w/U_f - 1$ vs turbulent intensity $U_{rms}/U_f$ for $\Theta = 7$ and $R = 5R_c$ (solid line). The dashed lines A, B show respective evaluation of flame velocity for a 3D case with and without the DL instability, respectively.  The markers show the experimental results (Aldredge et al, 1998).